\documentclass[prc,twocolumn,aps,showpacs]{revtex4}
\usepackage{amsmath,bm}
\usepackage{graphicx}
\begin{document}

\title{Isoscaling of projectile-like fragments}

\author{C. Zhong}\thanks{Present address: Brookhaven National
Laboratory, Upton, New York 11973, USA}

\author{Y. G. Ma} \thanks{Corresponding author. Email: ygma@sinap.ac.cn}

\author{D. Q. Fang}
\author{X. Z. Cai}
\author{J. G. Chen}
\author{W. Q. Shen}
\author{W. D. Tian}
\author{K. Wang}
\author{Y. B. Wei}

\author{J. H. Chen}
\author{W. Guo}
\author{C. W. Ma}
\author{G. L. Ma}
\author{Q. M. Su}
\author{T. Z. Yan}
\author{J. X. Zuo}
\affiliation{Shanghai Institute of Applied Physics, Chinese
Academy of Sciences, P.O. Box 800-204, Shanghai 201800, China}

\date{\today}

\begin{abstract}
The isotopic and isotonic distributions of the projectile
fragmentation products have been simulated by a modified
statistical abrasion-ablation (SAA) model and the isoscaling
behavior of projectile-like fragments has been discussed. The
isoscaling parameters $\alpha$ and $\beta$ for hot fragments
before evaporation and cold fragments after evaporation have been
extracted, respectively. It looks that the evaporation has strong
effect on $\alpha$. For cold fragments, a monotonic increase of
$\alpha$ and $|\beta|$ with the increasing of  $Z$ and $N$ is
observed. The relation between isoscaling parameter and the change
of isospin content is demonstrated. In addition, the disappearance
of the isospin effect of projectile fragmentation is also
discussed in the viewpoint of isoscaling parameter.
\end{abstract}
\pacs{ 25.70.Mn, 42.10.Pa}

\maketitle \nopagebreak

Recently, isoscaling behavior  has been extensively observed
\cite{Tsang2001PRL,Tsang2001PRC1,Tsang2001PRC2,Botvina}. The
scaling law relates ratios of isotope yields measured in two
different nuclear reactions, 1 and 2,
$R_{21}(N,Z)=Y_2(N,Z)/Y_1(N,Z)$. In multifragmentation events,
such ratios are shown to obey an exponential dependence on the
neutron number $N$ or proton number $Z$ of the isotopes
characterized by three parameters $\alpha$, $\beta$ and C
\cite{Tsang2001PRL}:
\begin{equation}
   R_{21}(N,Z) = \frac{Y_2(N,Z)}{Y_1(N,Z)} = C exp(\alpha N + \beta
    Z).
\end{equation}
In grand-canonical limit, $\alpha = \Delta\mu_{n}/T$ and $\beta =
\Delta\mu_{z}/T$ where $\Delta\mu_{n}$ and $\Delta\mu_{z}$ are the
differences between the neutron and proton chemical potentials for
two reactions, respectively. $C$ is an overall normalization
constant. This behavior is attributed to the difference of isospin
asymmetry between two reaction systems in the similar nuclear
temperature. It is potential to probe the isospin dependent
nuclear equation of state \cite{DiToro,Ma_review} by the studies
of isoscaling
\cite{Tsang2001PRL,Tsang2001PRC1,Tsang2001PRC2,Botvina,Ma2004PRC,Tian}.
So far, the isoscaling behavior has been studied  experimentally
and theoretically for different reaction mechanisms. However, most
studies focus on the isoscaling behaviors for light particles. A
few studies on the heavy projectile-like residues in deep elastic
collisions and fission fragments have been reported
\cite{Souliotis2003PRC,Friedman,Veselsky2,Wang,Ma_fis}. In this
paper, we concentrate our attentions on the isoscaling features in
projectile fragmentation and discuss the effect of evaporation on
the isoscaling parameters $\alpha$ and $\beta$  in a framework of
statistical abrasion-ablation (SAA) model \cite{Brohm1994NPA}.

Projectile fragmentation is one of the most important methods to
produce  beams of extreme neutron-rich or proton-rich nuclei and
has been widely used to study nuclear reactions induced by heavy
ions at intermediate and high energies. Various physical models
for projectile fragmentation
 have been developed and the reaction mechanisms of heavy-ion
 collisions have been investigated extensively.
 For instance, an empirical parameterization,
  like EPAX/EPAX II formula presented by S\"ummerer \emph{et al.} \cite{Suemmerer1999,Suemmerer2000},
  can predict the fragment production cross-section well.
   However, due to its poor physical foundation, the
  empirical parameterization may not be applicable in extrapolation.
In addition, statistical abrasion-ablation  model can describe the
isotopic distribution well \cite{Brohm1994NPA}. In the SAA model ,
the nuclear reaction is described in two stages which occur in two
distinctly different time scales. The first abrasion stage is
fragmentation reaction which describes the production of the
prefragment with certain amount excitation energy through the
independent nucleon-nucleon collisions in the overlap zone of the
colliding nuclei.  The collisions are described by a picture of
interacting tubes. Assuming a binomial distribution for the
absorbed projectile neutrons and protons in the interaction of a
specific pair of tubes, the distributions of the total abraded
neutrons and protons are determined. For an infinitesimal tube in
the projectile, the transmission probabilities for neutrons
(protons) at a given impact parameter $b$ are calculated by
\cite{Brohm1994NPA}
\begin{equation}
t_{\mbox{k}}(r-b)=\exp\{-[D_{\mbox{n}}^{\mbox{T}}(r-b)\sigma_{\mbox{nk}}+%
D_{\mbox{p}}^{\mbox{T}}(r-b)\sigma_{\mbox{pk}}]\},
\end{equation}
where $D^{\mbox{T}}$ is thickness function of the target, which is
normalized by $\int d^2rD^{\mbox{T}}_{\mbox{n}}=N^{\mbox{T}}$ and
$\int d^2rD^{\mbox{T}}_{\mbox{p}}=Z^{\mbox{T}}$ with
$N^{\mbox{T}}$ and $Z^{\mbox{T}}$ referring to the neutron and
proton number in the target respectively, the vectors $r$ and $b$
are defined in the plane perpendicular to beam, and
$\sigma_{\mbox{k}'\mbox{k}}$ is the free nucleon-nucleon cross
sections (k$'$, k$=$n for neutron and k$'$, k$=$p for proton). The
thickness function of the target is given by
\begin{equation}
D^{\mbox{T}}_{\mbox{k}}(r)=\int_{-\infty}^{+\infty}dz\rho_{\mbox{k}}
((r^2+z^2)^{1/2}),
\end{equation}
with $\rho_{\mbox{k}}$ being the neutron (proton) density
distribution of the target. So the average abraded mass at a given
impact parameter $b$ is calculated by the expression
\begin{equation}
\begin{array}{ll}
\langle \Delta A(b) \rangle= & \int d^2rD_{\mbox{n}}^{\mbox{P}}(r)[1-t_{\mbox{n}}(r-b)] \\
&+\int d^2rD_{\mbox{p}}^{\mbox{P}}(r)[1-t_{\mbox{p}}(r-b)].
\end{array}
\end{equation}
The excitation energy of projectile spectator is estimated by a
simple relation of $E^* = 13.3 \langle \Delta A(b) \rangle$ MeV
where 13.3 is a mean excitation energy due to  an abraded nucleon
from the initial projectile \cite{Gimmard}. In the second
evaporation stage the system reorganizes due to excitation, which
means it deexcites and thermalizes by the cascade evaporation of
light particles. By inducing in-medium nucleon-nucleon cross
section and optimizing computational method proposed by our group
given in Ref.
\cite{Cai2002PRC,Fang2001CPL,Fang2000PRC,Zhong2003HEP}, it can
give a good agreement with the isotopic distribution at both high
and intermediate energies involving neutron-rich or proton-rich
nuclei over a wide energy
range\cite{Fang2001CPL,Fang2000PRC,Zhong2003HEP}. The isospin
effect and its disappearance in projectile fragmentation for
$^{36,40}Ar$ at intermediate energies have been predicted by this
model and conformed by experimental data\cite{Fang2000PRC}.

\begin{figure}[t]
\includegraphics[width=7cm]{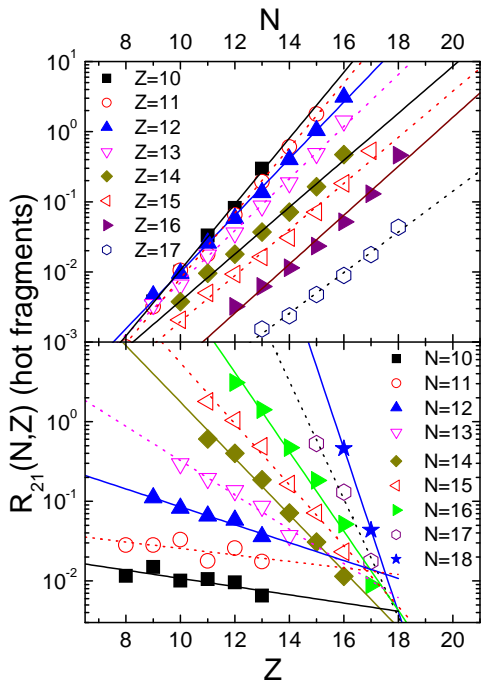} 
\caption{Yield ratios $R_{21}(N,Z)$ of pre-fragments (also called
hot fragments in this work) from the reactions of $^{40}Ar+^9Be$
and $^{36}Ar+^9Be$ at 60 MeV/nucleon versus  $N$ (top panel) or
$Z$ (bottom panel). The lines represent the exponential fits.}
\label{Pre-R21_Proton_Neutron}
\end{figure}

\begin{figure}[t]
\vspace{0.2cm}
\includegraphics[width=7cm]{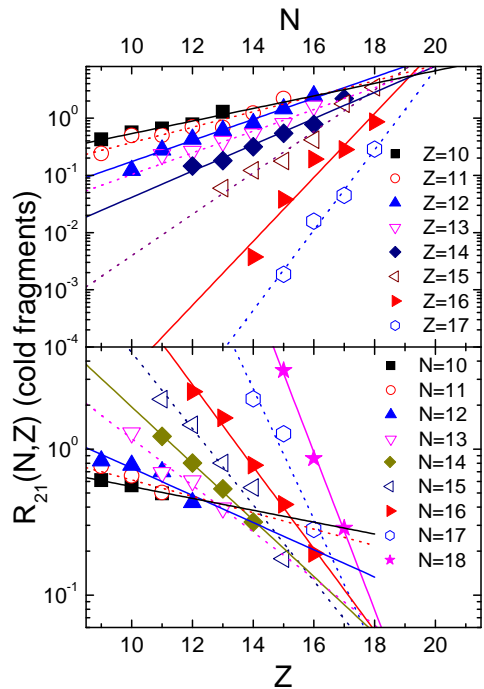} 
\caption{Yield ratios $R_{21}(N,Z)$ of cold fragments from the
reactions of $^{40}Ar+^9Be$ and $^{36}Ar+^9Be$ at 60 MeV/nucleon
versus  $N$ (top panel) or $Z$ (bottom panel). The lines represent
the exponential fits.} \label{R21_Proton_Neutron}
\end{figure}

In order to study the isoscaling effect in projectile
fragmentation, the reactions of $^{40/36}Ar + ^9Be$ at 60
MeV/nucleon were simulated by SAA model. We extract the yield
ratio $R_{21}(N,Z)$ using the convention that index 2 refers to
the more neutron-rich system ($^{40}Ar + ^9Be$) and index 1 to the
less neutron-rich one ($^{36}Ar + ^9Be$). Figure
\ref{Pre-R21_Proton_Neutron} shows the yield ratios $R_{21}(N,Z)$
of  hot projectile-like fragments (PLFs) as a function of neutron
number N for selected isotopes (upper panel) and proton number Z
for selected isotones (lower panel) for the $^{40/36}Ar + ^9Be$
reactions in the semi-log plots. In figure
\ref{R21_Proton_Neutron}, the corresponding ratios for the cold
projectile-like fragments are shown in the semi-log plots. In the
figures, the different isotopes and isotones are shown by
alternating filled and open symbols: even isotopes and isotones
are shown with filled symbols while odd ones are shown with open
symbols.

\begin{figure}[t]
\includegraphics[width=7cm]{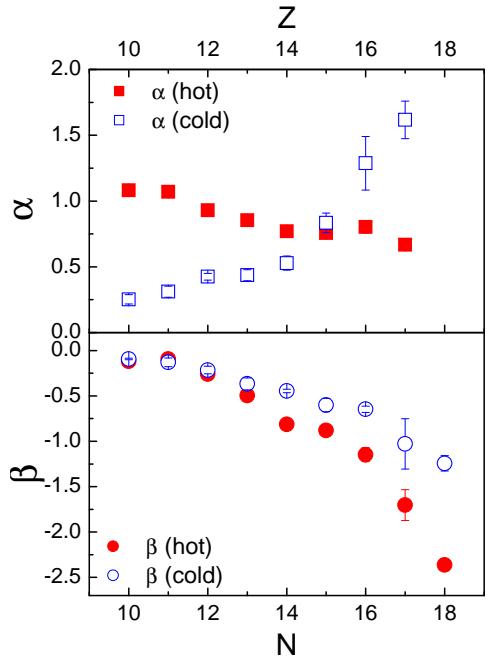} 
\vspace{.cm} \caption{Isoscaling parameters $\alpha$ as a function
of $Z$ (top panel) and $\beta$ as a function of $N$ (lower panel)
for projectile fragmentation from the reactions $^{40}Ar+^9Be$ and
$^{36}Ar+^9Be$ at 60 MeV/nucleon. Solid symbols represent the
pre-fragments and open symbols represent final fragments. }
\label{alpha_beta}
\end{figure}

From the upper panels of Fig.~\ref{Pre-R21_Proton_Neutron} (hot
fragments) and Fig.~\ref{R21_Proton_Neutron} (cold fragments), we
observe that the ratio for each isotope $Z$ (from 10 to 17)
exhibits a remarkable exponential behavior. For each isotope
($Z$), an exponential function form $C \exp(\alpha N)$ was used to
fit the calculation points  and the parameters $\alpha$ are shown
in the top panel of Fig. \ref{alpha_beta} for the selected $Z$.
Analogous behavior is observed in the lower panels of Figs.
\ref{Pre-R21_Proton_Neutron} and \ref{R21_Proton_Neutron}, for
each isotone ($N$), an exponential function  form $C' \exp(\beta
Z)$ was used to fit the calculation points and the parameters
$\beta$ are shown in the lower panel of Fig.~\ref{alpha_beta} for
the selected $N$.

The model predicts that the fragment isotopic distribution  at a
fixed atomic number $Z$ shifts towards the neutron-rich side for
the neutron-rich projectile. In Fig. \ref{alpha_beta}, we present
the slope parameters $\alpha$ (upper panel) and $|\beta|$ (lower
panel) of the exponential fits obtained as described for Figs.
\ref{Pre-R21_Proton_Neutron} and \ref{R21_Proton_Neutron} as a
function of $Z$ and $N$, respectively. In  the upper panel of Fig.
\ref{alpha_beta}, $\alpha$ of the pre-fragments for each element
show a little bit decreasing trend with the increasing of $Z$.
But, a monotonic increase of $|\beta|$ with the  increasing of $N$
is observed. For cold fragments which are survived after
evaporation (open symbols of Fig.~\ref{alpha_beta}), a monotonic
increase of $\alpha$ and $|\beta|$ with the increasing of $Z$ and
$N$, respectively, is observed. From the lower panel of
Fig.~\ref{alpha_beta}, $|\beta|$ always increase with the
increasing of $N$ for hot fragments or cold fragments and
$|\beta|$ of hot fragments tends to drop when evaporation is taken
into account.

However, some cautions should be reminded. Since  the first
abrasion stage is essentially a fast geometrical abrasion stage,
no equilibrium can be expected for the pre-fragments. These
fragments are only intermediate stage products and will decay due
to their excitation. In the strict sense, there is no isoscaling
behavior for this pre-fragments due to the lack of equilibrium.
However, a similar analysis as the isoscaling can be done since
there exists different isotopic/isotonic distributions between two
systems. In this work, this kind of analysis is useful to
investigate the evaporation effect on the isoscaling parameters of
PLF fragments \cite{Tsang2001PRC1}.

In the evaporation stage, neutrons are emitted with the most
important probability, the final cold fragments are more symmetric
than their hot ancestors, which reflects a strong evaporation
effect on projectile-like fragments \cite{Goldhaber,Ma_RC}.
However, the charged particle emission is less important. Since
$\alpha$ is the isoscaling parameter in fixed $Z$ and $\beta$ is
the isoscaling parameter in fixed $N$, so the neutron evaporation
has the strongest effect on $\alpha$ while charged particle
evaporation has less effect on $\beta$. That results in an obvious
change of $\alpha$ parameter while a weak change of $\beta$ due to
the evaporation.

\begin{figure}[t]
\includegraphics[width=7cm]{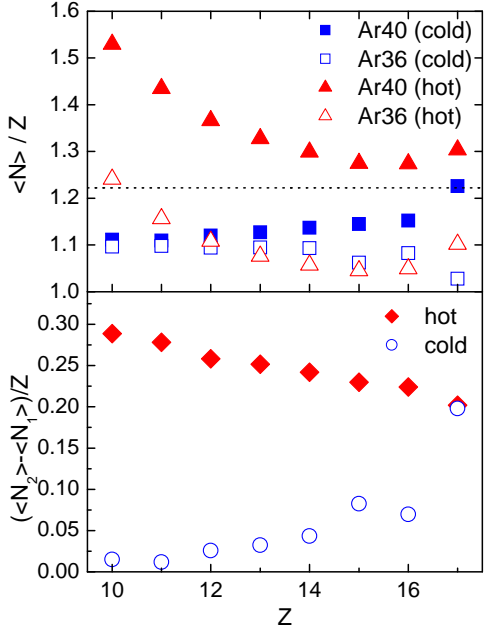} 
 \caption{(a) $\langle N\rangle/Z$ for hot and cold fragments
 for  $^{40}Ar$ and $^{36}Ar$ systems. The dotted line is initial N/Z for
  $^{40}Ar$; (b) The differences of $\langle N\rangle/Z$
   between  $^{40}Ar$ and $^{36}Ar$ for
the hot and cold fragments systems. The symbols are illustrated in
insert.} \label{nz}
\end{figure}

In order to further discuss the reason of the dramatic change of
the  isoscaling parameters, we plot $\langle N \rangle/Z$ versus
$Z$ in the upper panel of Fig.~\ref{nz} for the fragments before
and after evaporation. As known, $\langle N \rangle/Z$ represents
the isospin component. From the figure, the hot  PLFs which are
close to initial projectile basically remain larger isospin
content as the initial projectile while lighter hot PLFs have much
larger $\langle N \rangle/Z$ due to stronger dissipation and
nucleon exchange. However, this high isospin content can not keep
alive since the prefragments are excited as illustrated the above
section. They will cool themselves by the light particle
evaporation, mostly by the neutron emission. After the evaporation
process, $\langle N \rangle/Z$ tends to 1.1 in lower $Z$ and
bifurcates in higher $Z$ for $^{40}Ar$ and $^{36}Ar$ systems. More
interestingly, we note that the differences of $\langle N
\rangle/Z$ between two systems for hot and cold fragments as a
function of $Z$ as shown in the lower panel of Fig.~\ref{nz}.
Consistently, this behavior is similar to $\alpha$ as a function
of $Z$ for hot and cold PLFs as shown in Fig.~\ref{alpha_beta}. In
this context, we could say the difference of isospin content (i.e.
the neutron-to-proton ratio: $\langle N \rangle/Z$) is a more
simple measurement of isoscaling parameter $\alpha$. Similar
behavior has been found for $|\beta|$.

In the past years, the isospin dependence of various physical
quantities has been reported
\cite{Miller1999,Dempsey,DiToro,Muller1995,Ma1999,Fang2000PRC}.
The positive slopes in the upper panels of Figs.
\ref{Pre-R21_Proton_Neutron} and \ref{R21_Proton_Neutron} indicate
that neutron-rich fragments are more efficiently produced, as
expected, from the more neutron-rich systems. In other words, the
fragment isotopic distribution at a fixed atomic number $Z$ shifts
towards the neutron rich side for the neutron-rich projectile.
This isospin effect will decrease with decreasing the fragment
atomic number $Z$ and disappear when $(Z_{proj}-Z)/Z_{proj}$
becomes larger \cite{Fang2000PRC}. The quantity of
$(Z_{proj}-Z)/Z_{proj}$ can be related to excitation energy or
temperature of the deformed projectile after the abrasion stage.
When this quantity becomes smaller, the deformed projectile
becomes hotter or highly excited. In this context, we can say that
the isospin effect tends to fade out with the increasing of the
excitation energy of the projectile-like system.

In summary, the isoscaling of projectile-like fragments from
$^{40/36}Ar+^{9}Be$ at 60 MeV/nucleon has been studied by a
modified statistical abrasion-ablation model. SAA model can
simulate well the isotopic distribution of PLFs of the systems.
The isoscaling parameters $\alpha$ and $\beta$ are extracted for
the prefragments and final fragments. $\alpha$ has a dramatic
change after evaporation. For hot PLFs, $\alpha$ shows a little
bit stronger isospin effect for those highly excited small PLFs.
However, this kind of stronger isospin effect for smaller PLFs can
not survive due to much more neutron evaporation from these highly
excited pre-PLFs. Finally $\alpha$ shows weaker for these smaller
PLFs while stronger for PLFs close to initial projectile. A
monotonic increase of $\alpha$ ( $|\beta|$ ) with increase $Z$ (
$N$ ) is observed. This behavior can be easily explained by the
difference of the isospin content of hot or cold PLFs between two
systems as shown in Fig.~\ref{nz}. In this sense, isospin content
can take a simple "isometre" as isoscaling parameter. In addition,
the disappearance of the isospin effect of projectile
fragmentation is also discussed in the viewpoint of isoscaling
parameter.

This work was supported in part by the Shanghai Development
Foundation for Science and Technology under Grant Numbers
05XD14021 and 03 QA 14066, the National Natural Science Foundation
of China under Grant No 10328259, 10135030, 10405032, 10405033,
10475108, the Major State Basic Research Development Program under
Contract No G200077404.

{}

\end{document}